\documentclass[aps,twocolumn]{revtex4}
\usepackage{graphicx}
\begin{document}
\newtheorem{theorem}{Theorem}
\newtheorem{acknowledgement}[theorem]{Acknowledgement}
\newtheorem{algorithm}[theorem]{Algorithm}
\newtheorem{axiom}[theorem]{Axiom}
\newtheorem{claim}[theorem]{Claim}
\newtheorem{conclusion}[theorem]{Conclusion}
\newtheorem{condition}[theorem]{Condition}
\newtheorem{conjecture}[theorem]{Conjecture}
\newtheorem{corollary}[theorem]{Corollary}
\newtheorem{criterion}[theorem]{Criterion}
\newtheorem{definition}[theorem]{Definition}
\newtheorem{example}[theorem]{Example}
\newtheorem{exercise}[theorem]{Exercise}
\newtheorem{lemma}[theorem]{Lemma}
\newtheorem{notation}[theorem]{Notation}
\newtheorem{problem}[theorem]{Problem}
\newtheorem{proposition}[theorem]{Proposition}
\newtheorem{remark}[theorem]{Remark}
\newtheorem{solution}[theorem]{Solution}
\newtheorem{summary}[theorem]{Summary}    
\def\r{{\bf{r}}}
\def\i{{\bf{i}}}
\def\j{{\bf{j}}}
\def\m{{\bf{m}}}
\def\k{{\bf{k}}}
\def\kt{{\tilde{\k}}}
\def\mt{{\hat{t}}}
\def\mG{{\hat{G}}}
\def\mg{{\hat{g}}}
\def\mGa{{\hat{\Gamma}}}
\def\mS{{\hat{\Sigma}}}
\def\mT{{\hat{T}}}
\def\K{{\bf{K}}}
\def\P{{\bf{P}}}
\def\q{{\bf{q}}}
\def\Q{{\bf{Q}}}
\def\p{{\bf{p}}}
\def\x{{\bf{x}}}
\def\X{{\bf{X}}}
\def\Y{{\bf{Y}}}
\def\F{{\bf{F}}}
\def\G{{\bf{G}}}
\def\bG{{\bar{G}}}
\def\mbG{{\hat{\bar{G}}}}
\def\M{{\bf{M}}}
\def\V{\cal V}
\def\tchi{\tilde{\chi}}
\def\tx{\tilde{\bf{x}}}
\def\tk{\tilde{\bf{k}}}
\def\tK{\tilde{\bf{K}}}
\def\tq{\tilde{\bf{q}}}
\def\tQ{\tilde{\bf{Q}}}
\def\si{\sigma}
\def\ep{\epsilon}
\def\hep{{\hat{\epsilon}}}
\def\al{\alpha}
\def\be{\beta}
\def\ep{\epsilon}
\def\bep{\bar{\epsilon}_\K}
\def\up{\uparrow}
\def\de{\delta}
\def\De{\Delta}
\def\up{\uparrow}
\def\dwn{\downarrow}
\def\ksi{\xi}
\def\etha{\eta}
\def\product{\prod}
\def\goto{\rightarrow}
\def\switch{\leftrightarrow}
                           
\title{Magnetism in semiconductors: A dynamical mean field study of ferromagnetism in Ga$_{1-x}$Mn$_x$As }
\author{
K. Aryanpour$^{1,3}$, J. Moreno$^{2,3}$, M. Jarrell$^{3}$ and R.~S. Fishman$^{4}$ 
}
\affiliation{$^{1}$Department of Physics, University of California, Davis, California 95616}
\affiliation{$^{2}$Department of Physics and Astronomy, Clemson University, Clemson, South Carolina 29634}
\affiliation{$^{3}$Department of Physics, University of Cincinnati, Cincinnati, Ohio 45221}
\affiliation{$^{4}$Condensed Matter Sciences Division, Oak Ridge National Laboratory, Oak Ridge, 
Tennessee 37831}
\date{\today}

\begin{abstract}
We employ the dynamical mean field approximation to perform a systematic 
study of magnetism in Ga$_{1-x}$Mn$_x$As.  Our model incorporates the 
effects of the strong spin-orbit coupling on the $J=3/2$ GaAs valence bands and 
of the exchange interaction between the randomly distributed magnetic ions and the itinerant 
holes.  The ferromagnetic phase transition temperature $T_c$ is obtained for 
different values of the impurity-hole coupling $J_c$ and of the hole concentration 
$n_h$ at the Mn doping of $x=0.05$. We also investigate the temperature 
dependence of the local  magnetization and spin polarization of the holes.   
By comparing our results with those for a single band Hamiltonian in which the  
spin-orbit coupling is switched off, we conclude that the spin-orbit 
coupling in Ga$_{1-x}$Mn$_x$As gives rise to frustration in the ferromagnetic 
order, strengthening recent findings by Zar\'and and Jank\'o (Phys.\ Rev.\ Lett.\ 
{\bf{89}}, 047201 (2002)).

\end{abstract}

\maketitle  
            
\par 
The discovery of ferromagnetism in GaAs doped with Mn has renewed interest in 
the properties of diluted magnetic semiconductors.\cite{reviews}  Since these 
materials are good sources of polarized holes, they may form the basis of a 
new field called {\it spintronics}.\cite{Wolf01}
Spintronic devices employ both the spin and the charge of the carrier to 
convey information. 

\par In Ga$_{1-x}$Mn$_x$As, the Mn ions are in the Mn$^{2+}$ state with a 
half-filled $d$ shell of total spin $S=5/2$.\cite{Linnarsson97,Okabayashi98} 
Since Mn$^{2+}$ ions primarily replace Ga$^{3+}$, they act as effective acceptors 
by supplying holes as well as localized spins.  The valence band of pure 
GaAs is $p$-like so the strong spin-orbit interaction couples the  $l=1$ angular 
momentum of the $p$ orbitals to the electron spin ($s=1/2$), resulting in a 
total spin $J=l+s=3/2$ for the valence holes.\cite{blakemore} 
As discussed by Zar\'and and Jank\'o,\cite{janko} the strong spin-orbit
coupling also induces an {\it anisotropic} carrier-mediated interaction between
the Mn ions and, as a consequence, frustration in their ferromagnetic order.
While the results of Ref.~\onlinecite{janko}
were limited to the metallic regime and to small values of the Mn-hole 
coupling, a different approach suggests the presence of an impurity band
in the dilute limit.\cite{Zarand03}

In this Letter, we employ the Dynamical Mean Field Approximation 
(DMFA)\cite{metzner-vollhardt,muller-hartmann,pruschke,georges} to perform a 
systematic analysis of ferromagnetism in Ga$_{1-x}$Mn$_x$As, including the 
effects of strong spin-orbit coupling on the $J=3/2$ GaAs valence band.  The DMFA includes 
the spin-split impurity band through quantum self-energy corrections which are 
not included in other mean-field theories.  Because this method is
non-perturbative, it allows us to study both the metallic 
and impurity-band regimes as well as both small and large couplings. We show
how the spin-orbit interaction affects the ferromagnetic critical 
temperature $T_c$, the hole polarization, and the Mn magnetization.  
By comparing our results with those for a single band 
Hamiltonian without spin-orbit coupling, we conclude that strong 
spin-orbit coupling in Ga$_{1-x}$Mn$_x$As produces frustration for all 
coupling strengths. For carrier concentrations smaller than
the doping, both $T_c$ and the polarization of the carriers are reduced 
for all values of the coupling. For larger carrier densities
and large couplings, we unexpectedly find that frustration induces
a small but finite $T_c$, 
in sharp contrast with the vanishing $T_c$ found when the spin-orbit is neglected.

Our starting point is the simplified Hamiltonian proposed in Ref.~\onlinecite{janko}:
\begin{equation}
\label{eq:JanZand-nchiral-hamlt}
H=H_0-J_{c}\sum_{R_i}{\bf S}_i\cdot\hat{\bf J}(R_i).
\end{equation}
The first term includes the electronic dispersion and the spin-orbit 
coupling of the $J=3/2$ valence holes within the spherical approximation.\cite{Baldereschi73}
The second term is the dominant part of the interaction between the Mn spins 
and the valence holes,\cite{kacman} with $J_c$ the exchange coupling and 
$\hat{\bf J}(R_i)$ the total $J=3/2$ spin density of the holes at the site $i$ 
of a Mn ion with spin ${\bf S}_i$.  The relatively large magnitude 
of the Mn local moment justifies a classical treatment of its spin. 

Within the spherical approximation, the Hamiltonian of pure GaAs is 
rotationally invariant. Hence, $H_0$ is diagonal in a {\it chiral} basis, 
\begin{equation}
\label{eq:JanZand-chiral-sp}
H_0=\sum_{\k,\gamma}\frac{k^2}{2m_{\gamma}}{\tilde{c}^{\dag}_{\k\gamma}}
\tilde{c}_{\k\gamma}\,,
\end{equation}
where $\tilde{c}^{\dag}_{\k,\gamma}$ creates a chiral hole with momentum $\k$ 
parallel to its spin and $\hat{\bf J}\cdot\hat{\k}=\pm 3/2$ or $\pm 1/2$. The two 
effective masses $m_{\gamma}=m_h\approx0.5m$ and $m_{\gamma}=m_l\approx0.07m$ 
correspond to the heavy and light bands with $\gamma=\pm 3/2$ and $\pm 1/2$ 
respectively ($m$ is the electron mass). 

Unfortunately, the exchange interaction term obtains a rather complicated 
momentum-dependent form in the chiral basis\cite{janko} that is responsible 
for the frustrated order of the Mn. The competition 
between the strong spin-orbit coupling on the hole bands, which 
aligns the hole spin parallel to its momentum,  and the exchange 
interaction with the local moment, which aligns the hole and local spins,
precludes all of the carrier density from mediating the magnetic order.

\par We develop a DMFA algorithm that takes advantage of the simple diagonal 
form of $H_0$ in the chiral basis and the local form of the exchange interaction
in the non-chiral basis.  The coarse-grained Green function matrix in the 
non-chiral fermion basis is
\begin{equation}
\label{eq:non-chiral-gf}
\hat{G}(i\omega_n)=\frac{1}{N}\sum_{\k}[i\omega_n\hat{I}-\hat{\epsilon}(k)+\mu \hat{I}-
\hat{\Sigma}(i\omega_n)]^{-1}\,,
\end{equation}
where $N$ is the number of $\k$ points in the first Brillouin zone and 
$\displaystyle \hat{\epsilon}(k)= 
\hat{R}^{\dag}(\hat{\k}) \frac{k^2}{2m_{\gamma}}\hat{R}(\hat{\k})\,$
is the dispersion in the spherical approximation. Here,  $\hat{R}$ are spin 
$3/2$ rotation matrices that relate the fermion operator, $c_{\k\gamma}$
to its chiral  counterpart, 
$\tilde{c}_{\k\gamma}=R_{\gamma \nu}(\hat{\k}) c_{\k \nu}$. 
The mean-field function 
$\hat{\cal{G}}(i\omega_n)=[\hat{G}^{-1}(i\omega_n)+\hat{\Sigma}(i\omega_n)]^{-1}$
is required to solve the DMFA impurity problem.
At a nonmagnetic site, the local Green function equals the mean-field 
function, $\hat{G}_{non}=\hat{\cal{G}}$. At a magnetic site, $\hat{G}_{mg}$ must solve
a local problem. By treating disorder in a fashion similar 
to the {\it coherent potential approximation} (CPA)\cite{taylor} for a 
given local spin configuration, we obtain 
$\hat{G}_{mg}(i\omega_n)=[\hat{\cal{G}}^{-1}(i\omega_n)+J_{c}{\bf S}\cdot\hat{\bf J}]^{-1}$. 

Now $\hat{G}_{mg}(i\omega_n)$ must be averaged over all possible spin 
orientations at the local site and over all possible impurity configurations on the 
lattice. The former is implemented by using the effective action \cite{furukawa}
\begin{eqnarray}
\label{eq:eff-action}
S_{eff}({\bf S})=-\sum_{n}\log\det \left[ \hat{\cal{G}}(i\omega_n)
(\hat{\cal{G}}^{-1}(i\omega_n)+J_{c}{\bf S} \cdot\hat{\bf J})\right]
\nonumber\\&&\hspace*{-8.0cm}\times 
e^{i\omega_n0^+}\,
\end{eqnarray}
to average over the angular distribution of the local spins:
\begin{equation}
\label{eq:gf-or-avrg}
\left\langle\hat{G}_{mg}(i\omega_n)\right\rangle=\frac{1}{\cal Z}\int d\Omega_{\bf S} 
~\hat{G}_{mg}(i\omega_n)\exp[-S_{eff}({\bf S})]\,,
\end{equation}
where
${\cal Z}=\int d\Omega_{\bf S}\exp[-S_{eff}({\bf S})]$. 
If the Mn ions are randomly distributed with probability $x$, 
the configurationally-averaged Green function reads
$\displaystyle \hat{G}_{avg}(i\omega_n)=\left\langle\hat{G}_{mg}(i\omega_n)\right\rangle x+
\hat{\cal{G}}(i\omega_n)(1-x)$.

Finally, the magnetization of the Mn ions can be calculated as: 
\begin{equation}
\label{eq:Mn-magntzn}
M^z=\frac{1}{\cal Z}\int d\Omega_{\bf S} S^z
\exp[-(S_{eff}({\bf S})-\beta {\bf S}\cdot {\bf \delta H}^z)]\,,
\end{equation}
where a small magnetic field ${\bf\delta H}^z$ is applied to break the 
symmetry along a preferential direction, i.e., the $z$ axis.\cite{magfield}
By fitting the magnetization $M^z(T)$ in the vicinity 
of the transition to the Curie-Weiss form, we can extract the value of $T_c$ for 
each set of parameters studied.

We focus on the doping of $x=0.05$ associated with the highest $T_c$ 
reported.\cite{mbe,Ohno92,Ohno96,VanEsch97,Ohno98} In order to clearly 
elucidate the role played by the spin-orbit coupling, we introduce two  
parameters. One is the hole mass of an equivalent system composed of two
degenerate bands with the same average kinetic energy as our system:
$\displaystyle 2 m_{eq}^{\frac{3}{2}}={m_h}^{\frac{3}{2}}+{m_l}^{\frac{3}{2}}$.
The other is the ratio of light and heavy hole effective masses, 
$\displaystyle \alpha=\frac{m_l}{m_h}\,.$
The chirality can be switched off by setting $\alpha=1$ while keeping $m_{eq}$ 
constant. 

\begin{figure}
\includegraphics[width=3.4in]{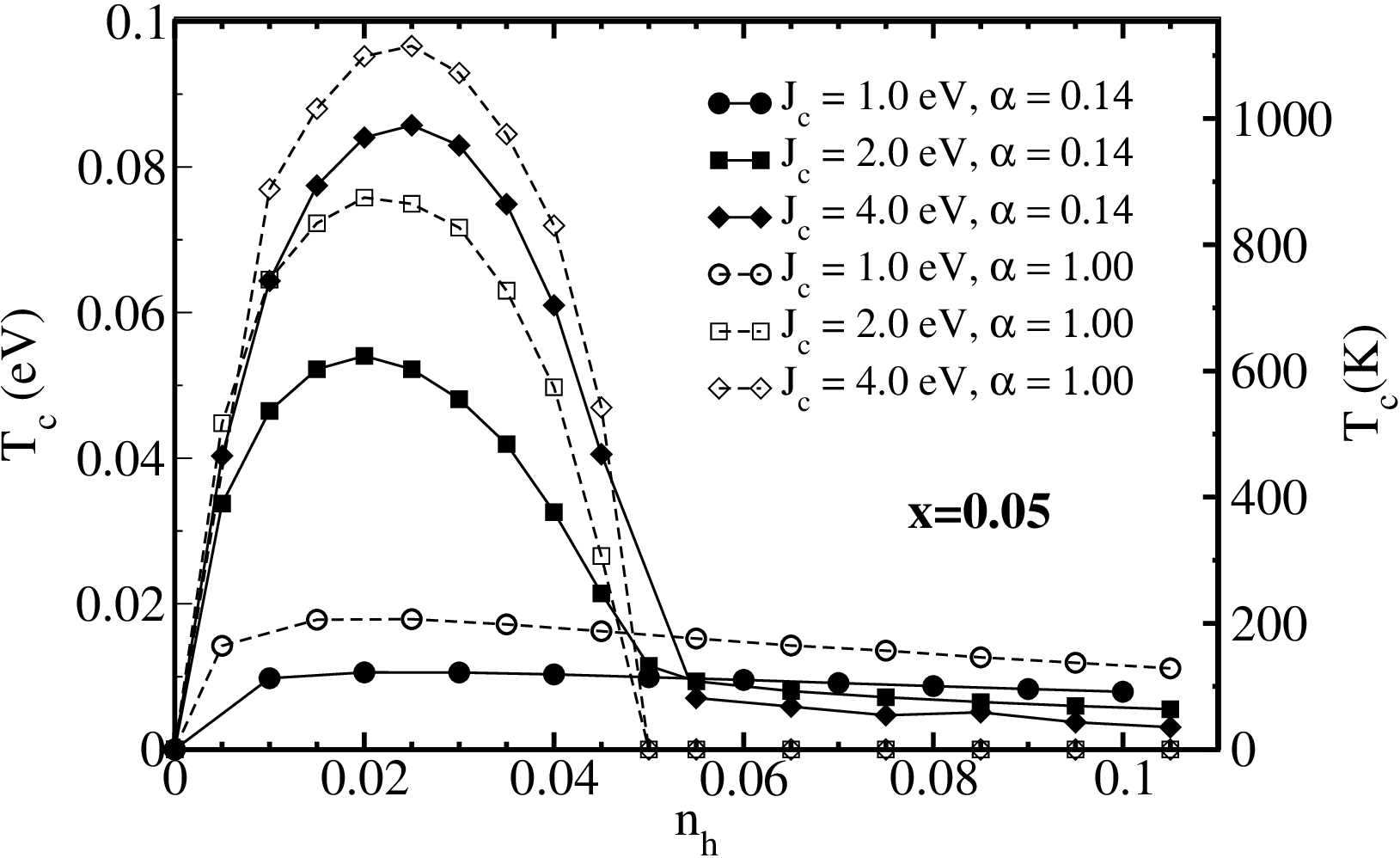}
\caption[a]{Phase transition temperature $T_c$ versus carrier concentration 
$n_h$ for Mn doping $x=0.05$, various $J_c$ values and effective mass 
ratios, $\alpha$.} 
\label{Tc.vs.n}
\end{figure}

\par Fig.~\ref{Tc.vs.n} presents our results for the phase transition 
temperature $T_c$ as a function of the hole concentration $n_h$ for a fixed
Mn concentration, $x=0.05$, and for various values of $J_c$ and $\alpha$. 
There are two regimes corresponding to 
small and large values of the interaction strength, $J_c$.  For large 
$J_c$, the holes form bound states with the Mn impurities, an impurity band 
develops inside the GaAs gap, and the properties of the host are greatly 
affected.  The value of $J_c$ at which the impurity
band appears depends upon the value of $\alpha$.

First consider the non-chiral case with $\alpha=1.0$ so that the spin-orbit 
coupling is turned off. For $J_c=1.0$ eV the system is far from the 
impurity-band regime and $T_c$ has a relatively slow variation with respect 
to $n_h$. For $J_c=2.0$ eV the system is beyond the threshold 
for the formation of an impurity band, which then dominates the physics.  The 
maximum $T_c$ occurs when the impurity band is nearly half filled 
($n_h\approx x/2$).\cite{chattopadhyay} For $J_c=4.0$ eV  the 
impurity band is well established and the maximum $T_c$ is large. Because
ferromagnetic order restricts the hopping of holes when the impurity band 
is full, $T_c$ vanishes for $J_c> 1.0$ eV and $n_h>x$. 
An antiferromagnetic ground state is energetically more favorable in 
the regime $n_h > x$ \cite{chattopadhyay} because the carriers can then 
easily hop from one impurity site to another.
Magnetization curves for $J=2.0$ and 4.0 eV and for $n_h > x$ provide 
evidence for antiferromagnetism.  They fit a Curie-Weiss functional 
form, $\displaystyle M^z= \frac{M^z_0}{1 + T/\Theta }$
with a positive $\Theta $ ranging from $40$ to $117 K$,
indicating antiferromagnetic order for $\alpha=1.0$ and $n_h> x$.

We now consider the chiral case with $\alpha=0.14$ so that the spin-orbit 
coupling in GaAs is turned on. The effects of frustration on the ferromagnetic 
critical temperature are readily seen in Fig.~\ref{Tc.vs.n}. For $J_c=1.0$ eV, 
$T_c$ consistently lies below its non-chiral  counterpart.  At $J_c=2.0$ and 
$4.0$ eV, the difference between the chiral and non-chiral results is even 
larger. For $n_h < x$, the non-chiral $T_c$ is always higher than the chiral $T_c$.
However, for $n_h > x$, the chiral results continue to yield 
a finite although greatly diminished $T_c$ in the same regime where 
$T_c$ vanished when $\alpha=1.0$.  
Surprisingly, the ferromagnetic $T_c$ survives at large dopings 
due to the intrinsic frustration in the system.  Strong spin-orbit 
scattering allows the impurity-band carriers to hop between impurity 
sites that are ferromagnetically aligned even when the impurity band is full, 
thereby stabilizing the magnetic order. 

\begin{figure}
\includegraphics[width=3.0in]{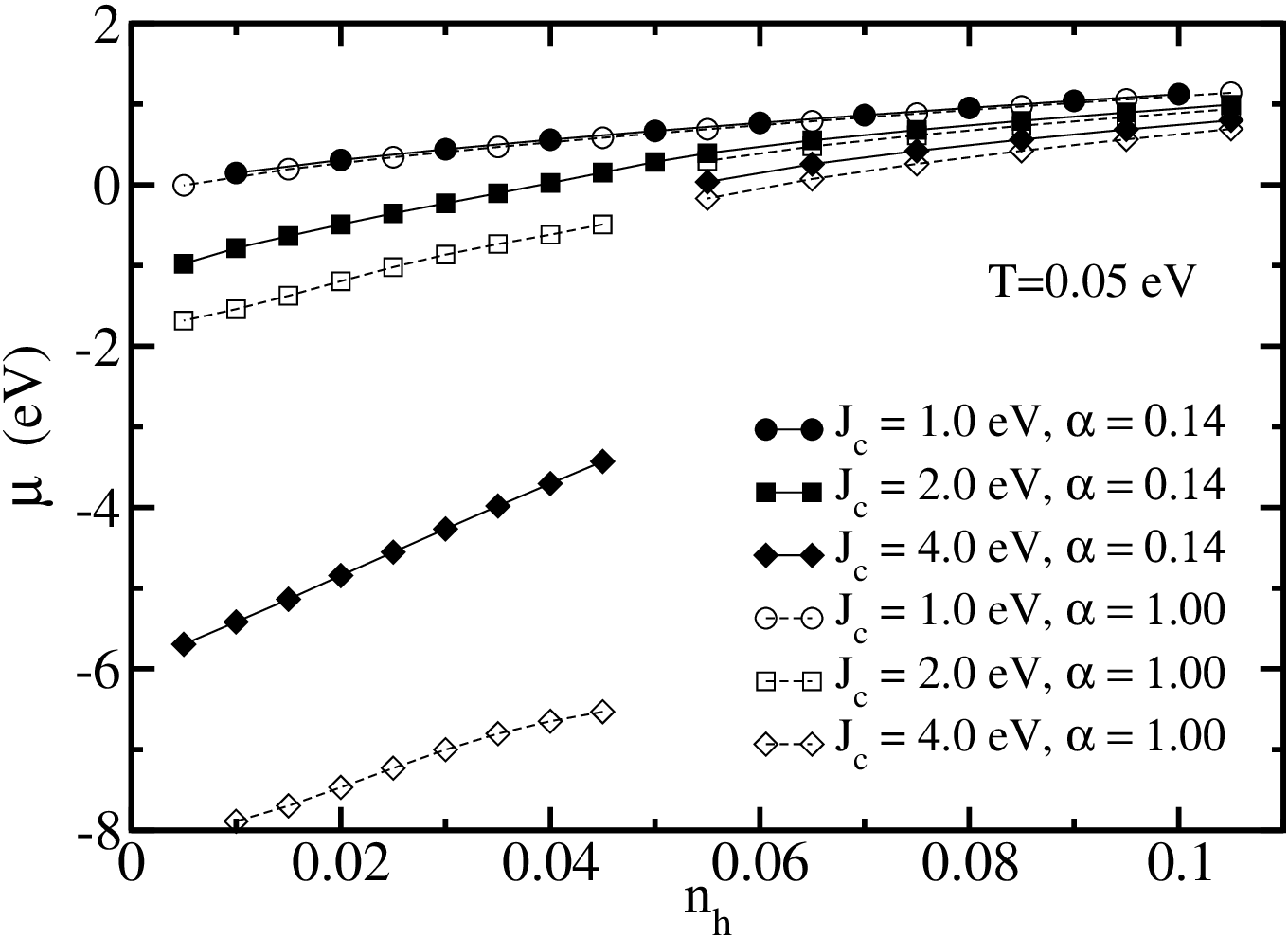}
\caption[a]{Chemical potential, $\mu$, versus $n_h$ for  Mn doping $x=0.05$ and
$T=0.05$ eV using the same values of $J_c$ 
and $\alpha$ as in Fig. 1.}
\label{mue.vs.n}
\end{figure}

The formation of an impurity band is responsible for many of the outstanding
properties of this system.\cite{chattopadhyay}  
Impurity-band formation appears as a discontinuity of 
the hole chemical potential $\mu$ at $n_h= x$.  Fig.~\ref{mue.vs.n} 
depicts $\mu$ versus $n_h$ for $T=0.05$ eV ($580 K$) and for the same values 
of $J_c$ and $\alpha$ used in Fig.~\ref{Tc.vs.n}. The presence of an impurity band 
depends on both $J_c$ and $\alpha$.  
When $J_c=1.0$, no impurity band is present 
so the chiral and non-chiral $\mu$ are almost identical and neither shows a 
discontinuity.  When $J_c=2.0$ eV and 
$\alpha=1.0$, the impurity band has already split from 
the valence band.  At the same value of $J_c=2.0$ eV but when 
$\alpha=0.14$, the change in slope of $\mu$ at $n_h=x$ 
indicates that the impurity band is present but that it overlaps with the main band.  
When $J_c=4.0$ eV and $\alpha=0.14$, the hole chemical potential is 
discontinuous at $n_h=x$, signaling the splitting of the impurity band from 
the valence band.  For these larger values of $J_c$, the non-chiral $\mu$ 
always lies below the chiral value, indicating that the non-chiral impurity 
band lies at lower energies than the chiral one.  Since the interaction between 
the impurity moments is mediated by the host of holes, the less pronounced 
impurity band for $\alpha=0.14$ is another signature of the frustration produced by
chirality.
  
Optical conductivity measurements\cite{Singley02} show that the impurity band 
is split from the valence band at the doping of $x=0.05$.  This fact and our results for 
$T_c$ suggests a value of  $J_c$ between the value of $1.2$ eV obtained from
photoemission techniques\cite{Okabayashi98} and the value of $4.5 \sim 6$ eV 
inferred from infrared spectroscopy.\cite{Linnarsson97} 

\par
\begin{figure}
\includegraphics[width=3.0in]{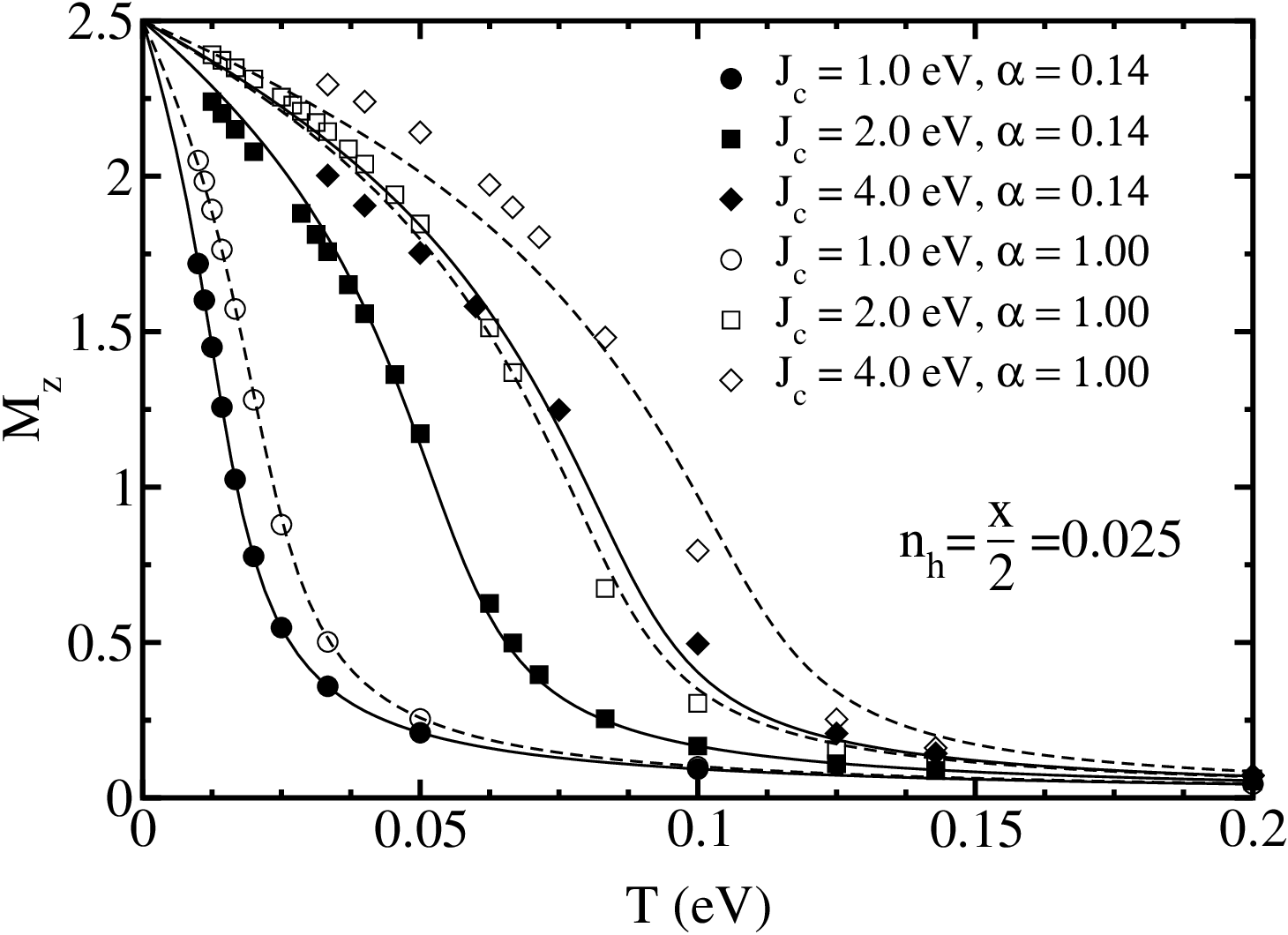}
\caption[a]{Magnetization versus temperature for different 
chirality modes and different 
values of $J_c$ at $n_h=x/2=0.025$ and $\delta H^z=0.004$ eV.  
The curves are the static mean field fits with $T_c$ 
values equal to those given by our results, solid curves fit the results for  $\alpha=0.14$;
dashed curves the ones for $\alpha=1$.}
\label{Imp.vs.T.n0.025}
\end{figure}

 The magnetization also reveals the effects of frustration.  In 
Fig.~\ref{Imp.vs.T.n0.025}, the temperature dependence of the magnetization
is plotted with
$n_h=x/2=0.025$. Our results are compared with Curie-Weiss fits for the 
same $T_c$ values. When 
$J_c=1.0$ eV, the mean-field curves perfectly fit the data  not only at large 
temperatures but also below $T_c$. But for larger $J_c$ values and low 
temperatures, discrepancies appear between the static mean-field curves and the DMFA 
results. For $\alpha=1.0$, the magnetization in our model lies above 
the mean-field Heisenberg magnetization when $T < T_c$ since the scattering 
between the itinerant carriers and the localized spins becomes coherent at 
low temperatures, thereby enhancing the magnetic order of the local ions.  This effect 
has also been seen in the double exchange model.\cite{Furukawa-private} 
However, for $\alpha=0.14$, frustration reduces the low-temperature coherence 
between the carriers and local ions.  So when $T < T_c$, the magnetization curves 
lie below the mean-field Heisenberg predictions.

\par
\begin{figure}
\includegraphics[width=3.0in]{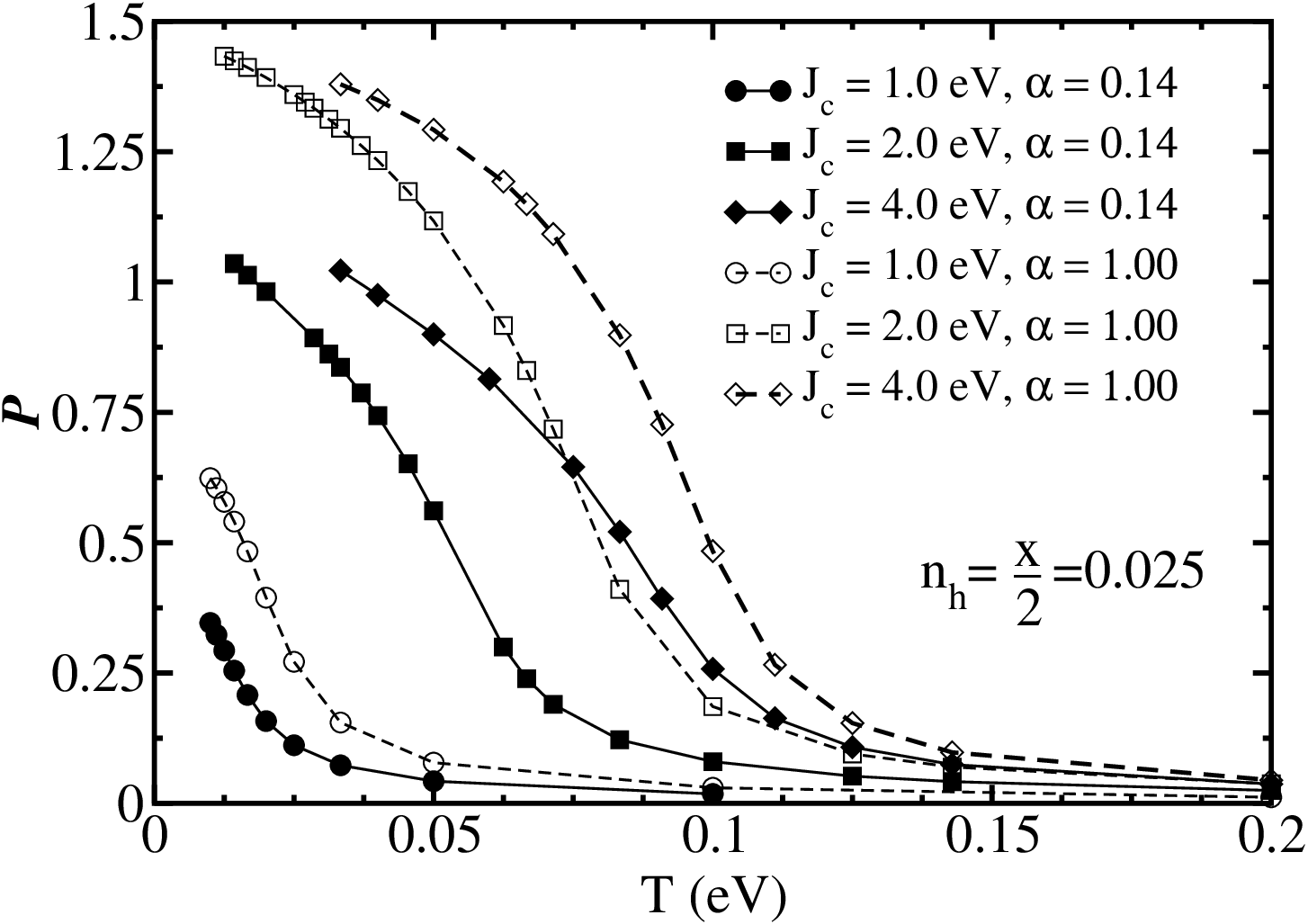}
\caption{The hole polarization {\cal P} versus temperature at $n_h=x/2=0.025$ and
${\delta H}^z=0.004$ eV.} 
\label{polaz.vs.T}
\end{figure}

 Lastly, we study the carrier spin polarization, ${\cal P}$, which is 
constructed from appropriate sums of the Green function. Control of ${\cal P}$
is important for {\it spintronics} applications, since the spin polarization 
of the holes is required to transport information.
Fig.~\ref{polaz.vs.T} illustrates the temperature dependence of the 
hole polarization when $n_h=x/2=0.025$ for several values of $J_c$ and $\alpha$. 
For all $J_c$ values, chirality suppresses the polarization. For $\alpha=1.0$, 
large $J_c$, and low temperatures, ${\cal P}$ 
approaches $3/2$ since all holes occupy the lowest 
energy level with $j_z=+3/2$.  But due to frustration even 
for large $J_c$ and low temperatures, ${\cal P}$  for $\alpha =0.14$
is significantly smaller.  This agrees with previous calculations 
of the zero-temperature polarization, where the destructive effects 
of the spin-orbit interaction were also found.\cite{Dietl01}

In summary, we have studied the diluted magnetic semiconductor Ga$_{1-x}$Mn$_x$As 
using a non-perturbative multi-orbital DMFA algorithm incorporating the effects
of the strong spin-orbit coupling.  
We have calculated the ferromagnetic critical temperature, 
hole chemical potential, local ion magnetization, and hole polarization for 
a broad range of model parameters.  We find that the spin-orbit coupling 
leads to frustration and reduced magnetization when the hole concentration 
$n_h$ is smaller than the impurity concentration $x$, in agreement with
previous perturbative calculations.\cite{janko}  In addition, we find that 
this behavior persists for large values of $J_c$, and that frustration greatly 
reduces the transition temperature $T_c$ and the polarization of the carriers 
for all $J_c$.  Finally, when $J_c$ is large, we find the surprising result that
frustration induces a region of finite $T_c$ for $n_h>x$.

This approach has promising future ramifications.  
It can be extended to study other magnetic semiconductors and
realistic devices such as semiconducting heterostructures and quantum 
dot systems, which can be tailored to take full advantage of the intrinsic 
anisotropy of the ferromagnetic order.
More sophisticated approaches, such as 
the dynamical cluster approximation (DCA),\cite{Hettler00}
may be used to go 
beyond the single-site approximation and explore the cooperative and glassy 
effects of frustration, such as the reduction in the local magnetization at 
low temperatures. 

\par We acknowledge useful conversations with N. Furukawa and B. Jank\'o.
The hospitality of the Condensed Matter Sciences Division at Oak Ridge 
National Laboratory is gratefully acknowledged.  
This research was supported 
by NSF grants DMR-0073308 and DMR-0312680 and by the Department of Energy 
grants no. DE-FG02-01ER45897 and DE-FG03-03NA00071 (SSAAP program) and also
under contract DE-AC05-00OR22725 with Oak Ridge National Laboratory,
managed by UT-Battelle, LLC.


\begin{thebibliography}{18}
\bibitem{magfield} We have used two values of the magnetic field $\delta H^z=0.004$ and 
$0.0002$ eV.
\end{thebibliography}
\end{document}